\pdfoutput=1
\documentclass[twoside]{article}

\usepackage{lipsum} 

\usepackage[sc]{mathpazo} 
\usepackage[T1]{fontenc} 
\usepackage{microtype} 

\usepackage[hmarginratio=1:1,top=32mm,columnsep=20pt]{geometry} 
\usepackage{multicol} 
\usepackage[small,up,textfont=up]{caption} 
\usepackage{booktabs} 
\usepackage{float} 
\usepackage{hyperref} 

\usepackage{paralist} 

\usepackage{abstract} 

\newcommand{\alphalarge}{\mathrel{\scalebox{1.2}{$\alpha$}}} 

\usepackage{titlesec} 
\renewcommand\thesection{\Roman{section}} 
\renewcommand\thesubsection{\Roman{subsection}} 
\titleformat{\section}[block]{\large\scshape\centering}{\thesection.}{1em}{} 
\titleformat{\subsection}[block]{\large}{\thesubsection.}{1em}{} 

\usepackage{graphicx}
\usepackage{epstopdf}
\usepackage{epsfig}
\usepackage[]{natbib}

\title{\vspace{-15mm}\fontsize{24pt}{10pt}\selectfont\textbf{Estimation Error of Expected Shortfall}} 

\author{
\large
\textsc{Imre Kondor}\thanks{E-mail: kondor.imre@gmail.com}
\\[2mm]
\normalsize Parmenides Foundation, Pullach b. M\"unchen, Germany\\ 
\vspace{-5mm}
}
\date{}

\begin{document}

\maketitle 



\begin{abstract}

The problem of estimation error of Expected Shortfall is analyzed, with a view of its introduction as a global regulatory risk measure.

\end{abstract}



\section{Introduction}
\label{sec_int}

As part of the overhaul of market risk regulation, the Basel Committee on Banking Supervision advocates the introduction of Expected Shortfall (ES) to replace VaR as the standard regulatory risk measure (\citet{bas12,bas13}). The industry, of course, complains, but also various academic groups have expressed criticism concerning ES (e.g. \citet{zie03}, \citet{con10}, or \citet{dan13}).

One has to understand that a risk measure tries to grasp the risk in a typically very large portfolio, that is to compress the huge amount of information contained in a multivariate probability distribution into a single figure. There is no unique way to perform this compression such that it optimally fit every conceivable portfolio, therefore any risk measure will always be an easy target for criticism. One has to weigh several contradictory requirements and find a meaningful and practical compromise. I believe the Basel Committee made an eminently sensible choice by selecting ES.

The purpose of this note is to briefly mention the advantageous features of ES, to point to one of its obvious yet seldom advertised weaknesses, and to suggest regularization as a remedy. The assessment of the risk of a given portfolio and the selection of the optimal portfolio are obviously very different tasks, but both entail the element of out of sample projection, and if the career of VaR is anything to go by, ES will soon be promoted from a diagnostic tool also into a decision making tool. Therefore I discuss the two functions together.

\section{The merits of ES} 
\label{sec_merits}

Let me first recapitulate the merits of ES:

\begin{itemize}

\item ES is easy to conceptualize: if one is able to comprehend a quantile (VaR), one must be able to comprehend the average beyond that quantile (ES) too.
\item	ES characterizes the dangerous tail fluctuations much better than VaR, and is harder to manipulate. 
\item	ES is easy to measure, and can be optimized via linear programming (\citet{roc00}).
\item	ES is a coherent risk measure (\citet{ace01, ace02}) in the sense of \citet{art97, art99} and as such it is convex. I do not believe one should be addicted to axiomatic thinking in the context of such a complex topic as financial risk, but convexity seems to me a \textit{sine qua non} of any acceptable risk measure. A non-convex risk measure penalizes diversification, may lead to regulatory arbitrage, does not allow risk to be properly aggregated and priced (because it does not guarantee the efficient frontier to be convex), does not guarantee that a consistent nested system of limits can be constructed at large institutions, etc. In contrast, VaR can be concave, although \cite{dan13b} show that, except for extremely heavy tailed asset return distributions, VaR is sub-additive in the tail region. 

Let me add as an aside that the risk measures implied by the standard model for calculating the capital charge of various positions in the old market risk regulation (see e.g. \citet{cad98}) can also be concave (\citet{kon04}). Hopefully, similar errors will be avoided in the new regulation.

\end{itemize}

\section{High dimensionality} 
\label{sec_dimensionality}

The weakness I wish to bring into focus is not specific to ES, it stems from the large size of institutional portfolios, and from the relatively small amount of available information. Therefore, portfolio selection and risk management of large portfolios should be regarded as problems in high dimensional statistics (\citet{buh11}).
As such, they display huge estimation errors. Similar problems permeate a large number of fields beyond finance, and modern statistics has developed efficient tools called regularizers for their treatment. They come under colorful names like ridge regression, shrinkage, support vector machines, lasso, elastic net, etc., and offer a rich toolset to tackle the different types of difficulties arising in high dimensional statistics. Their systematic application to risk measurement and portfolio selection is highly desirable, however it has to be kept in mind that different regularizers work differently with different risk measures, and they must be chosen judiciously, with due attention paid to the nature of the portfolio in question. There are plenty of details still to be explored in this area, especially concerning ES, if it is to become the new global regulatory risk measure. (Factor models, GARCH filtering, Monte Carlo bootstrap, and several other procedures widely used in the financial industry can also be regarded as some sort of regularization procedures. Their efficiency should be weighed against that of the regularization methods borrowed from high-dimensional statistics.) 

Regularization methods have been put forward in the context of portfolio theory by several authors (e.g. \citet{job79}; \citet{jor86}; \citet{fro86}; \citet{led03, led04b, led04}; \citet{gol07}; \citet{bro09}; \citet{dem09}; \citet{sti10}; \citet{cac10}), but the full scope and potential of the idea has not been exploited.

\section{The problem of estimation error} 
\label{sec_esterror}

Let me start the exposition of the estimation error problem with the example of the variance as a risk measure. There, it is well-known that the covariance matrix becomes singular (develops zero eigenvalues) when the number $N$ of different assets in the portfolio (the dimension of the problem) becomes equal to, or larger than, the length $T$ of the time series. The essential parameter of the problem is the ratio $N/T$; for small values of this ratio (i.e. for long time series, a large amount of information available) the estimation error of the risk of a given portfolio, or the error in the selection of the optimal portfolio will be small. For $N/T$ approaching $1$ from below, the estimation error becomes large, and for $N/T = 1$ it actually diverges, grows beyond any finite value (\citet{kon07}). 

This is, of course, a simple linear algebraic problem, the number of unknowns becoming larger than the number of equations. It can, however, be viewed also as an algorithmic phase transition (\citet{mez09}). This way of looking at the problem has the advantage that it allows one to take over some key concepts from the theory of phase transitions. One of these is the concept of universality: around the critical point some characteristics of the transition become independent of the details of the cost function. For example, one can show that the estimation error is growing as the $-1/2$ power of the distance from the critical point $(N/T)_c = 1$. This exponent is universal, it does not depend on the character of the fluctuations of the portfolio elements, it is the same whether these fluctuations are Gaussian, fat tailed or even GARCH-like (\citet{var07}).

There are several ways how one can get rid of the nuisance of a singular covariance matrix, one of them is to add a constant to its diagonal elements, thereby preventing it from developing zero eigenvalues. This is a simple example of regularization and goes by the name of shrinkage (\citet{led03, led04b, led04}). One can choose many different priors to which to shrink the covariance matrix, each of them representing different bits of external information projected into the (biased) estimation. 

\begin{figure}[H] 
  \centering
  \includegraphics[width=0.6\columnwidth]{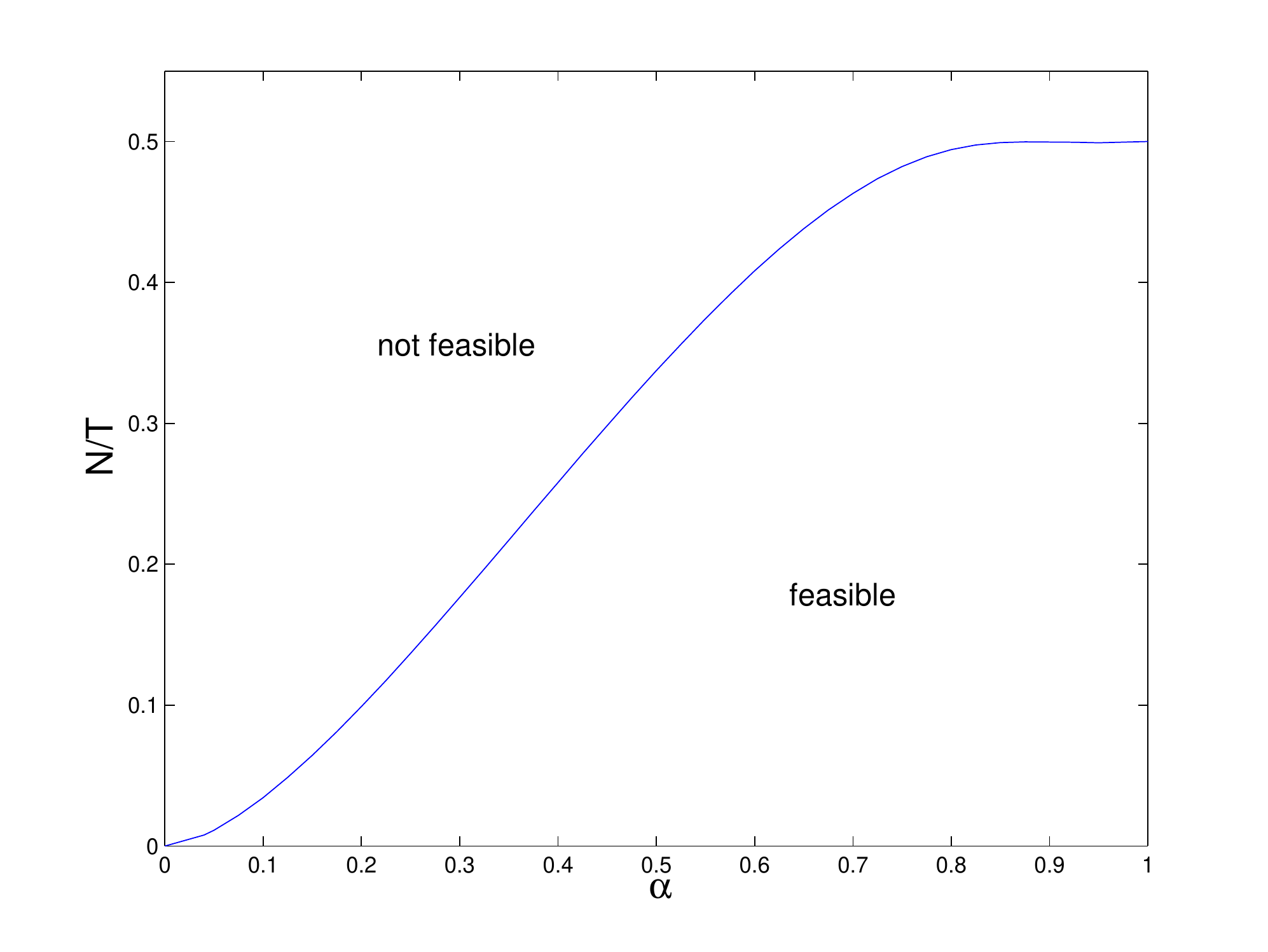}
  \caption{The phase diagram of ES: Above the transition line the optimization of ES is not feasible (ES is not bounded from below), while below the phase transition line the optimization problem always has a finite solution. Approaching the transition line from below the estimation error diverges as the $-1/2$ power of the distance from the line. }
 \label{psd_es}
\end{figure}

A similar difficulty is also present in the problem of ES. If we do not have enough information we will end up with large errors in the estimation of portfolio risk, while the optimization of the portfolio will become unfeasible. There is a difference from the case of the variance, however. While for the variance the instability sets in at $N=T$ and portfolio optimization is always possible (perhaps with a large error) for any $T>N$, for ES we can have samples where the optimization is unfeasible even for large $T$'s. The probability of these samples may be small for a given $N$ and $T$, but it is never exactly zero. For both $N$ and $T$ large, the probability of these exceptional samples becomes negligible, and we will have a sharp transition again, with a critical ratio $(N/T)_c$ that will now depend on the threshold $\alphalarge$ beyond which we calculate the average loss ES. For $\alphalarge\; = 1$ (that is the extremal case of ES where we consider not the worst $1\%$ or $5\%$, but \textit{the single worst} outcome and optimize this over the portfolio weights) the critical ratio will be $(N/T)_c = 1/2$ (\citet{kon07}). For any  other value of $\alphalarge$ the critical value of $N/T$ will be different, so we will have a critical line (a phase diagram) on the $N/T - \alphalarge$ plane. For Gaussian returns, this phase diagram is shown in Fig.~\ref{psd_es} (\citet{cil07}). For other underlying fluctuations the phase diagram may be different, may be shifted, etc. but upon approaching the critical line the estimation error will blow up with the same exponent $-1/2$ as in the case of the variance.

What is the financial meaning of these singularities? In the case of the variance, the first zero eigenvalue enters when two rows (or a linear combination of some rows and another row) of the covariance matrix become proportional to each other, that is when the returns on two assets move strictly parallel with each other in time. In the case of ES, the root of the singular behavior is different: the optimization of ES becomes unfeasible (the risk measure becomes unbounded from below) when there is an asset, or a linear combination of assets, in the portfolio that dominates all the others, that is produces a larger return at each time point $1,2,\dots T$. This can happen for any $N/T$, though for $N/T$ below the transition line the probability of this is small. In (\citet{kon10}) we showed that the same instability appears, for the same reason, under any coherent risk measure, irrespective of the concrete form of the stochastic process that drives the returns. 

\begin{figure}[H] 
  \centering
  \includegraphics[width=0.6\columnwidth]{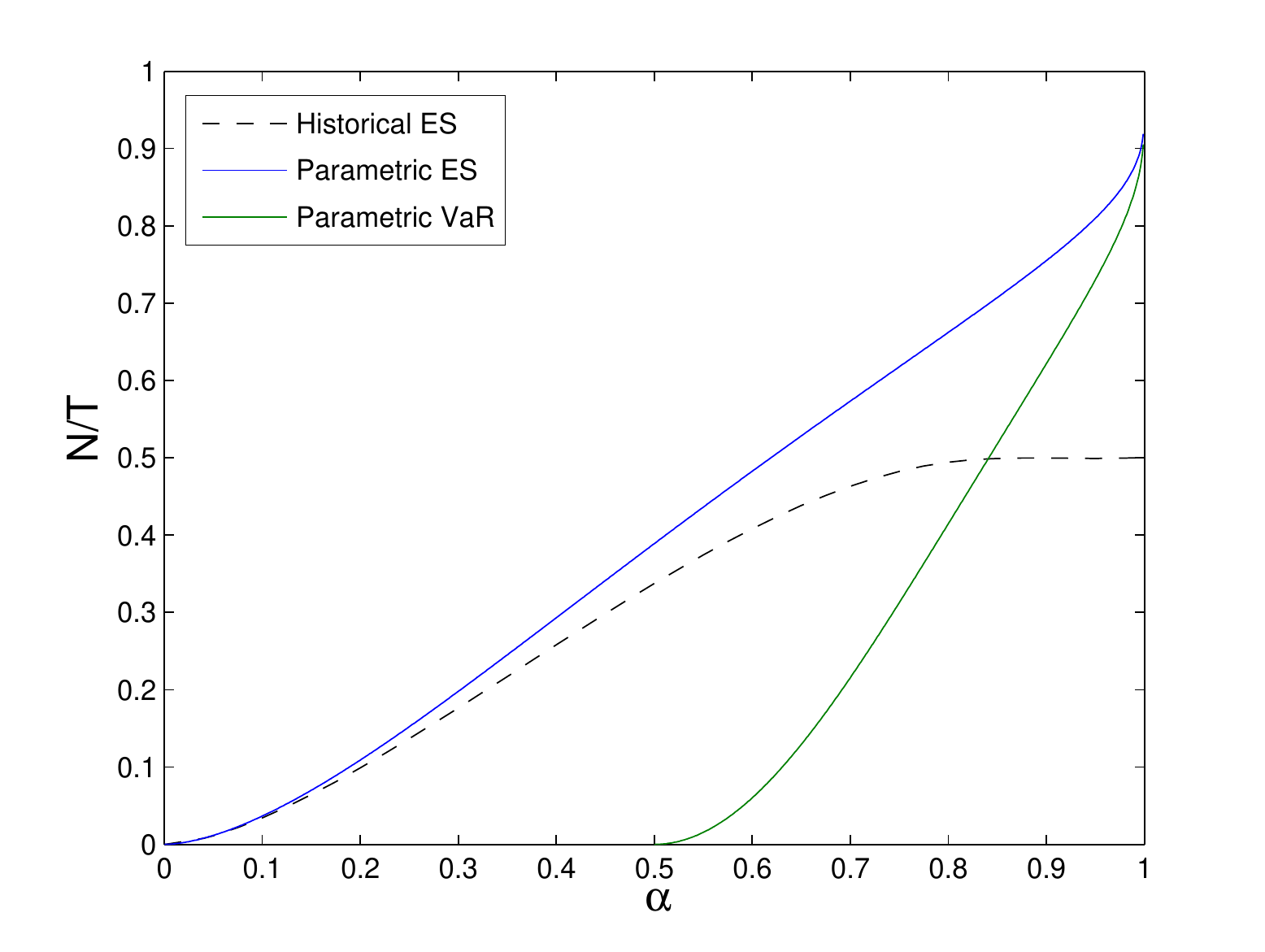}
  \caption{The phase transition line of historical ES (dashed line, the same as in Fig.~\ref{psd_es}), parametric ES (blue line) and parametric VaR (green line). The risk measures can be optimized below their respective transition lines, but are unbounded from below above these lines. The parametric ES and parametric VaR lines arrive at $\alpha = 1$ with an infinite slope; the critical value of both functions is $N/T = 1$ at this point. }
 \label{ptl_es}
\end{figure}

Now for a finite sample, the probability of finding a dominant asset will always be (perhaps small, but) finite, even if in reality, on longer time horizons, no such dominance relationship exists between the assets. In such a situation, the optimal choice for an investor who is guided by ES is to go very long (,,infinitely long'') in the dominant asset, and go correspondingly short in the rest. This way the expected shortfall (or any other coherent measure) can be made \textit{negative} and \textit{arbitrarily large} in absolute value. Negative risk means no risk, so when a dominant asset exists in the portfolio the investor can make an arbitrarily large profit at no risk at all. We can see then that ES (and all its coherent sisters) may give a false arbitrage alert with finite (though maybe small) probability on the basis of finite samples, even if there is no real arbitrage present.

It must be clear from the above that the difficulty is not related to the particular nature of ES or coherence in general, but to the fact that they are downside risk measures. Indeed, in (\citet{var08}) we showed that the same instability is present also in parametric VaR. (We analyzed parametric VaR because its historic counterpart can become concave and the methods we used cannot handle concave risk measures.) The phase diagrams of historic ES, parametric ES and parametric VaR for Gaussian returns are shown in Fig.~\ref{ptl_es}.

\section{Conclusion}
\label{sec_conclusion}

Downside risk measures were introduced because investors are not supposed to be worried about big wins, only about big losses. Perhaps they should be. Downside risk measure can create the mirage of fake arbitrage and lure investors into very large positions that implode when the mirage disappears in the next sample. The US housing bubble was a macroscopic example of such a folly.

Downside risk measures induce short-termism. If the whole industry is applying a risk measure that issues false arbitrage signals, this becomes a major element of systemic risk. It is therefore highly important that the Basel Committee construct a risk measurement rule that removes this undesirable feature of ES by stipulating that appropriate regularization procedures be introduced at banks. The concrete choice and implementation of the regularizer may be debatable but it must be chosen such that it does not instigate short term risk taking whereby to trigger herding behavior.

Clearly, no regularizer will ever replace human judgment, sober deliberation and ,,moral sentiment''. Selecting the proper risk management procedure is more than a problem in statistics or mathematics. It is also a problem in ethics. Resisting the lure of fake arbitrage and declining the short term advantage in favor of the long term interest and stability of business, these are moral qualities that must spread in the whole industry if we want to avoid a repeat of the madness of the previous decade and the string of scandals of recent years. Regulation will be successful only if it succeeds in promoting this deep reform of culture in finance.

\section*{Acknowledgment}

I am indebted to Istv\'an Csabai, Alan Kirman, Imre Risi Kondor, Ole Peters and G\'abor Papp for helpful remarks, and M\'at\'e Cs. S\'andor for assistance with the manuscript. This work has been supported by the European Union under grant agreement No. FP7-ICT-255987-FOC-II Project and by the Institute for New Economic Thinking under grant agreement ID: INO1200019.


\bibliographystyle{abbrvnat}

\bibliography{es}   
%
%
%



\end{document}